\documentclass[final,5p,times,authoryear]{elsarticle}

\usepackage{caption}

\usepackage{amssymb}
\usepackage{multirow}
\usepackage{amsmath}
\usepackage{booktabs}
\usepackage{subcaption} 

\journal{ISPRS Journal of Photogrammetry and Remote Sensing}

\begin{document}

\begin{frontmatter}

\title{High-Resolution Forest Mapping from L-Band Interferometric SAR Time Series using Deep Learning over Northern Spain}

\author[label1,label2]{Chiara Telli\corref{cor1}}
\ead{chiara.telli@uniroma1.it}

\author[label3]{Oleg Antropov\corref{cor1}}
\ead{oleg.antropov@vtt.fi}

\author[label3]{Anne Lönnqvist}
\author[label2]{Marco Lavalle}

\cortext[cor1]{Corresponding authors}

\address[label1]{Sapienza University, Dept. of Information Eng., Electronics and Telecommunications, 00184 Roma, Italy}
\address[label2]{Jet Propulsion Laboratory, California Institute of Technology, Pasadena, CA 91011, USA}
\address[label3]{VTT Technical Research Centre of Finland, Tekniikantie 1, 02150 Espoo, Finland}

\begin{abstract}
In this study, we examine the potential of high-resolution forest mapping using L-band interferometric time series datasets and deep learning modeling. Our SAR data are represented by a time series of nine ALOS-2 PALSAR-2 dual-pol SAR images acquired at near-zero spatial baseline over a study site in Asturias, Northern Spain. Reference data are collected using airborne laser scanning. We examine the performance of several candidate deep learning models from UNet-family with various combinations of input polarimetric and interferometric features. In addition to basic Vanilla UNet, attention reinforced UNet model with squeeze-excitation blocks (SeU-Net) and advanced UNet model with nested structure and skip pathways are used. Studied features include dual pol  interferometric observables additionally incorporating model-based derived measures. Results show that adding model-based inverted InSAR features or InSAR coherence layers  improves retrieval accuracy compared to using backscatter intensity only. Use of attention mechanisms and nested connection fusion provides better predictions than using Vanilla UNet or traditional machine learning methods. Forest height retrieval accuracies range between 3.1-3.8 m ($R^2$ = 0.45--0.55) at 20 m resolution when only intensity data are used, and improve to less than 2.8 m when both intensity and interferometric coherence features are included. At 40 m and 60 m resolution, retrieval performance further improves, primarily due to higher SNR in both the intensity and interferometric layers. When using intensity at 60 m resolution, best achieved RMSE is 2.2 m, while when using all suitable input features the achieved error is 1.95 m.  
We recommend this hybrid approach for L-band SAR retrievals also suitable for NISAR and future ROSE-L missions.

\end{abstract}

\begin{graphicalabstract}
\centering
    \resizebox{0.85\textwidth}{!}{
\includegraphics{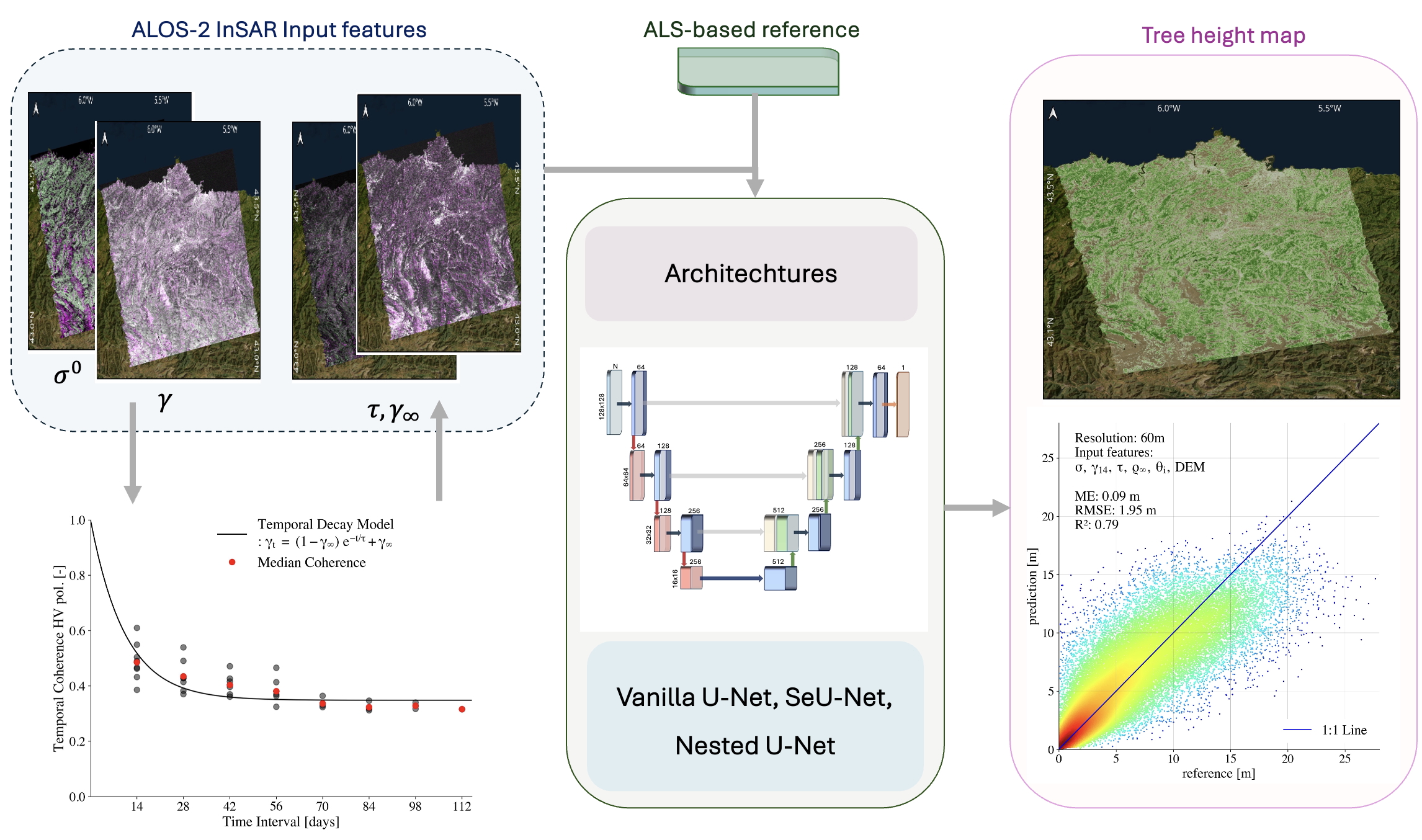}
}
\end{graphicalabstract}

\begin{highlights}

\item Demonstrated the ability of joint multi-temporal interferometric and backscatter observations from PALSAR-2 to capture structural variations in forest canopies.

\item Developed a hybrid physics--deep learning framework that combines InSAR coherence model--based inversion, polarimetric and auxiliary observables with UNet-family convolutional neural networks for forest height retrieval.

\item Achieved RMSEs between $\sim$ 2.8~m at 20~m resolution and 2.0~m at 60~m strongly outperform traditional machine learning or physics-based inversion approaches.
    
\item Attention mechanisms and nested feature fusion yield more spatially coherent forest height maps consistent with airborne laser scanning measurements.
    
\item Moderate spatial aggregation (40--60~m) provided an effective noise suppression enabling scalable monitoring of forest structure from satellite L-band InSAR imagery.
\end{highlights}


\end{frontmatter}


\section{Introduction}

\label{sec:intro}

Accurate methods for assessing forest carbon using Earth Observation (EO) sensors are increasingly needed to support climate change monitoring, carbon stock accounting, and to address the operational needs of diverse forest management stakeholders \citep{latterini2025}. Forests are essential in maintaining healthy ecosystem interactions and biodiversity on Earth, and they play a critical role in tackling climate change due to their capacity to rapidly restore carbon stocks. This ecological and climatic importance has placed forests at the center of international environmental policies, which call for accurate, timely, and reliable methods for large-scale forest monitoring \citep{berger2025,moersberger2024}.

At present, forest carbon stock and its changes are typically estimated through manual inventories of structural forest variables, a process that remains highly labor-intensive and limited in spatial and temporal coverage \citep{neumann2016, gfoi-2014, IPCC2003}.  

EO sensors offer opportunities to overcome these limitations by providing consistent, wall-to-wall coverage with regular temporal sampling.  The integration of satellite EO data with ground measurements has therefore become an established practice in forest mensuration and monitoring, enabling the estimation of forest variables across multiple spatial scales and mapping units \citep{mcrob2007}, and improving national forest inventories \citep{hunka2025}. Such estimates can be obtained either by inverting physics-based models, which aim to describe the interaction between transmitted signals and forest structure, or by applying data-driven approaches, which directly learn the underlying relationships from large sets of reference samples \citep{gfoi-2014}. 

Within this context, multiparametric satellite optical and synthetic aperture radar (SAR) imagery represent critical sources of information for large-scale forest mapping. Optical sensors provide valuable insights into vegetation phenology and health, but their applicability is often constrained by cloud cover and atmospheric conditions. In contrast, SAR observations, being independent of daylight and cloud cover, enable consistent monitoring across regions and seasons. The interaction between the radar signal and the observed scatterers is strongly influenced by the scatterers’ geometric and dielectric properties, making SAR particularly sensitive to key vegetation characteristics, such as forest height, growing stock volume, and above-ground  biomass \citep{ulaby2014, leToan2004} 

To date, extensive studies demonstrated significant correlations between both SAR backscatter coefficient and SAR interferometric measurements, and forest variables \citep{schmullius2015}. This sensitivity is influenced by several factors, including mission configuration, radar frequency, polarization, and the structural characteristics of the forest. Due to the enhanced penetration capability to vegetation structure compared with shorter wavelengths, L-band time series have proven particularly effective in estimating forest variables and are considered presently most suitable for SAR based forest variable retrieval \citep{gfoi-2014}. 
Currently reported retrieval accuracies, however, indicate that further improvements are needed to achieved more accurate and reliable estimates. The integration of deep learning techniques into EO-based forestry, together with the advent of new remote sensing missions offering advanced imaging capabilities, such as NISAR, holds strong potential for methodological advances when improved modeling approaches are combined with novel imaging systems. 
Research in this field remains active, with continued efforts to refine physically based retrieval models using L-band time series and to develop deep-learning frameworks for data-driven forest estimation parameters. The following sections summarize recent progress in both approaches, highlighting key trends and remaining challenges.

\subsection{L-band SAR time series in forest mapping}

SAR intensity based approaches infer forest structural variables from SAR backscatter time series by leveraging the sensitivity of low-frequency radar to forest structure using physics-based models and, more recently, machine learning techniques. Early research on L-band time series for forest mapping
focused primarily on estimating growing stock volume (GSV) in boreal forests using L-band SAR imagery, particularly from the JERS \citep{kurvonen1999,pulliainen1999,rauste2005,santoro2008}. Later studies utilized ALOS PALSAR and ALOS-2 PALSAR-2 sensors to map not only boreal forests \citep{cartus2012, antropov2013, antropov2017, santoro2024} but also temperate and tropical forests \citep{englhart2012,berninger2018,rodriguezveiga2019}, and even on global scales \citep{santoro2024cci} . 

While models based approaches based on Water Cloud Model \citep{santoro2024} provide a simplified representation of these processes, forward modeling of electromagnetic wave propagation within forested environments remains inherently complex due to multiple scattering interactions and the challenge of isolating contributions from distinct scattering components. In forested areas, L-band backscatter generally arises from three main components: (i) volume scattering from the canopy, (ii) surface scattering from the forest floor, particularly in canopy gaps, and (iii) double-bounce scattering caused by interactions between the ground and tree trunks. The sensitivity of radar backscatter intensity to forest height is further constrained by the well-known problem of signal saturation, which reduces the dynamic range of tree height estimates \citep{dobson1992dependence,mermoz2014biomass,mitchard2009using}. 

Potential improvement in estimating forest variables are possible by leveraging the interferometric potential of SAR acquisitions. Early and well-established approaches rely on the sensitivity of interferometric measurements to forest height, which arises from the angular separation between two SAR acquisitions \citep{treuhaft2000vertical}. A widely used physical model is the Random Volume over Ground (RVoG), which represents the forest scattering profile as the sum of a Dirac-like ground component and a vertical distribution of randomly oriented scatterers \citep{cloude1998polarimetric}. Inversion of the RVoG model typically requires a single-baseline, fully polarimetric (quad-pol) interferometric SAR acquisition, allowing the retrieval of forest height. However, most current and planned interferometric SAR missions, both at higher frequencies (e.g., C-band Sentinel-1) and at lower frequencies (e.g., L-band ALOS-2, NISAR, and ROSE-L), are designed to operate in dual-polarization modes with near-zero spatial baselines.  In these observation scenarios, sensitivity to vegetation structure due to the interferometric angular separation is reduced, and the influence of temporal vegetation dynamics on the interferometric phase and coherence becomes dominant. This has motivated recent efforts to retrieve forest parameters from time series of near-zero-baseline, dual-polarimetric interferometric observations at C- and L-band, where temporal effects play a major role \citep{lei2014,seppi2022,bhogapurapu2024, Cartus2022, lavalle2023}. 

For zero- or near-zero interferometric baseline configurations, interferometric measurements are mainly affected by temporal decorrelation, i.e., the decrease in the correlation coefficient between two radar complex images acquired at different times albeit with the same observation geometry. \cite{Lavalle2012} extended the RVoG model by introducing a vertical motion profile of canopy scatterers to represent wind-induced temporal decorrelation, as well as  dielectric variations that occur between acquisitions, linking these effects to forest structural properties. This model, known as the Random-Motion-over-Ground model (RMoG), has been used to describe temporal decorrelation mechanisms and to retrieve tree height from both Sentinel-1 C-band \citep{lavalle2023} and ALOS-2 L-band backscatter and interferometric coherence time series \citep{Telli2025}. 

A simplified approach to characterize temporal decorrelation is to model temporal coherence as an exponential decay \citep{sica2019repeat}. Although this behavior may not always hold at the pixel level, temporal coherence, particularly when considered in aggregated form, is generally expected to decrease over time due to the combined influence of multiple decorrelation sources. In this formulation, two parameters are defined: the \textit{temporal decay rate}, which describes how rapidly coherence decreases with increasing temporal baseline, and the \textit{long-term coherence}, which represents the residual coherence observed at very large time intervals. Together, these parameters absorb the effects of various decorrelation sources, including vegetation dynamics, soil moisture variability, and system noise. 

Given a time series of interferometric coherence maps, these two parameters can be estimated in a pixel-wise basis through curve fitting, thereby providing spatially explicit insights into the sensitivity of temporal coherence to forest structure. Compared with more complex models, this simplified approach requires fewer assumptions and less ancillary information, while still offering a practical way to exploit temporal decorrelation for forest monitoring. However, since the different sources of decorrelation are not explicitly separated, directly deriving forest variables from these parameters is not straightforward. This limitation motivates the deep learning approach proposed in this study, which aims to investigate whether hidden, non-linear relationships between temporal coherence dynamics and forest structure can be effectively captured to enable tree height estimation.

\subsection{Deep learning modeling and SAR data in forest mapping}

Deep learning (DL) methodologies are widely adopted in the remote sensing community for image classification and semantic segmentation tasks \citep{persello2022,zhu2020,zhu2017}. To date, several fully convolutional and recurrent neural network architectures were investigated for regression tasks in the context of EO-based forest inventory mapping \citep{astola2021,illarionova2022,ge2022improved,ge2022lstm, bolyn2022}. These models enable the integration of spatial and/or temporal context in addition to spectral or radiometric (polarimetric) observables within the modeling process. Thus, they typically achieve higher accuracy in forest classification, biophysical variable estimation(e.g., biomass, height, and volume), and forest change mapping \citep{kuzu2024} compared to traditional machine learning, statistical, or model-based approaches. Training these DL models generally requires a representative extensive dataset consisting of fully segmented reference labels, such as image patches derived from airborne laser scanning (ALS) digital surface models or ALS-based forest maps. Nevertheless, these models can be adapted to new sites through transfer learning strategies \citep{astola2021, ge_transfer2023}. 

To date, deep learning approaches in the context of forest mapping were studied using primarily Sentinel-1 images \citep{ge2022lstm}, sometimes augmented by optical image datasets \citep{ge2022improved}.

In addition, several studies explored the potential of deep learning for forest height mapping using TanDEM-X data. Although these studies reported excellent results with accuracies reaching 10-20\% RMSE \citep{mazza2019,carcereri2023deep}, the availability of TanDEM-X data remains limited, restricting its applicability for large-scale or operational forest monitoring.

In contrast, L-band time series have been investigated less frequently due to data availability constraints.  
In this study, we address this gap by utilizing a time series of PALSAR-2 interferometric acquisitions for forest mapping, leveraging advanced U-Net modeling approaches and a hybrid strategy that integrates model-based parameter retrieval with deep learning modeling.

\subsection{Our study goals and objectives}

The overarching objective of this work is to develop and evaluate a robust methodology for exploiting time series of interferometric L-band data dominated by temporal decorrelation to maximize the accuracy of forest height prediction, thereby supporting operational forestry practices and preparing for the uptake of NISAR mission data. The study relies on ALOS-2 PALSAR-2 datasets acquired over highly productive forests in Northern Spain, an area characterized by challenging topography and diverse forest structure.

For the deep learning framework, we rely on a family of U-Net models, which have consistently demonstrated high performance in predicting forest structural variables and are still competitive to the latest advances in geospatial foundation modeling. Three variants are considered: the classical U-Net \citep{ronneberger2015unet}, the nested U-Net \citep{zhou2018unet++}, and the Squeeze-and-Excitation U-Net (SeU-Net) \citep{ge2022improved, hu2020seunet}.

The specific objectives of this study are to:
\begin{enumerate}

    \item Evaluate the potential of advanced deep learning architectures, including nested and attention-enhanced U-Net variants, for quantitative forest-structure mapping from multi-temporal PALSAR-2 interferometric data, and benchmark their performance against conventional machine-learning and model-based approaches;

    \item Develop a hybrid UNet model based approach that combines input polarimetric observables and physics based inverted parameters within CNN modeling approach to improve forest height prediction performance;
       
    \item Quantify the contribution of different SAR observables to forest height retrieval accuracy within hybrid UNet model based prediction, and assess the added value of integrating physics-model derived parameters and geometric features such as incidence angle and DEM layers in forest height retrievals; and
    
    \item Examine the impact of spatial resolution and interferometric data processing on model accuracy and spatial fidelity, determining optimal trade-offs between detail and stability for PALSAR-2 based forest height retrieval. 

\end{enumerate}

To achieve these objectives, we combine dual-polarization ALOS-2 backscatter and interferometric time-series data with model-based features derived from a physics-based exponential decay model of temporal coherence \citep{sica2019repeat}, and UNet family based deep learning framework using airborne LiDAR measurements as reference. The approach includes a multi-resolution analysis to assess the robustness of the model across different spatial scales.

The paper is organized as follows: Section~\ref{sec:data} describes the study area and the datasets used in this work, Section~\ref{sec:methods} details the deep learning architectures and the overall methodological framework, and Section~\ref{sec:results} presents the experimental results and performance assessment. Finally, Section~\ref{sec:conclusion} summarizes the main findings and outlines directions for future research.

\section{Data}\label{sec:data}

\subsection{Study site}

\begin{figure}[t!]
    \centering
    \includegraphics[width=1.\linewidth]{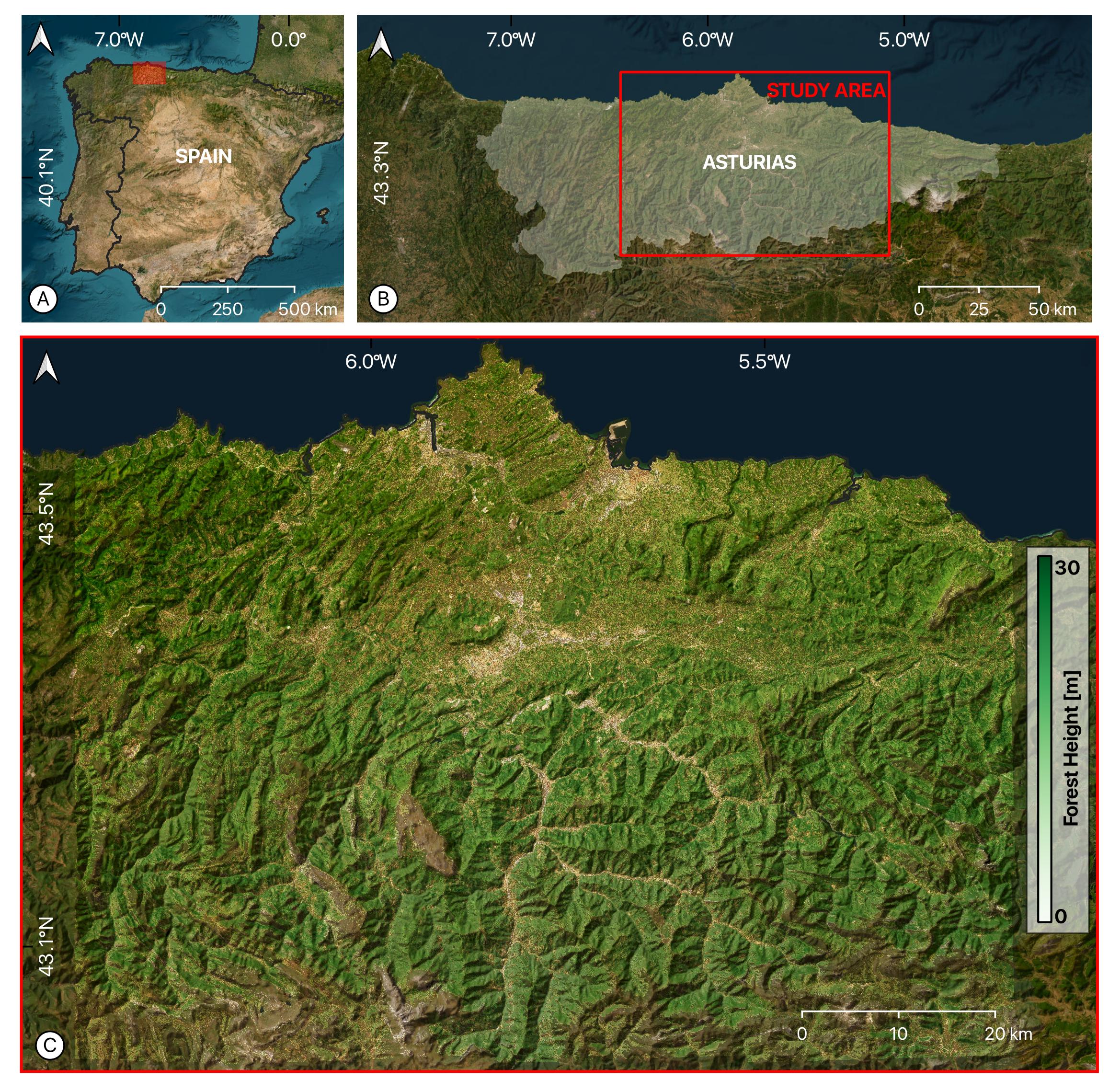}
    \caption{Location of the study site in Asturias, Spain. Panel B shows the extent of the study area within the Asturias region, and Panel C presents the detailed study area along with the reference tree height map  \citep{MDSnV2021}. Coordinate reference system: EPSG:4326, WGS84 (geodetic latitude/longitude).}
    \label{fig:study_site}
\end{figure}

The study area is located in northern Spain (longitude: -6.33°, -5.33°; latitude: 42.98°, 43.68°) within the Principality of Asturias. It is characterized by a Temperate Broadleaf and Mixed Forests biome, and it is predominantly covered by vegetation. In the area, the most common species are sweet chestnut (45.8\%), beech (25.9\%), blue gum (19.5\%), and oaks (4.5\%) \citep{MFE2023}. The site is characterized by rough topography, being relatively flat in the north and mountainous toward the south, with elevations ranging from 100 m to 2500 m. The climate is mild, with an average annual temperature of 10°C. The mean annual precipitation is approximately 1000 mm, with the highest rainfall occurring mainly during the winter months. Due to its proximity to the ocean, the region is frequently affected by strong winds. The study site is shown in Figure \ref{fig:study_site}.

\subsection{SAR and auxiliary data}
\label{subsec:sar_aux_data}
The SAR dataset consists of a time series of 9 dual-pol (HH, HV) ALOS-2 SLC images acquired at 14-day intervals between
June and September 2020. Data were collected in Fine Beam Dual (FBD) mode along an ascending orbit. The local acquisition time is approximately midnight, and the off-nadir angle is 28°.  

The ALOS-2 data processing was performed using JPL’s ISCE3 (InSAR Scientific Computing Environment) software \citep{rosen2018insar}. All SAR images were orthorectified and radiometrically normalized with respect to the projected incidence angle area to obtain gamma-naught ($\gamma^0$) images using the ISCE3 Area-Based Projection Algorithm \citep{shiroma2022area}, which averages the backscatter measurements using an adaptive multi-looking window. Orthorectification was performed using Copernicus DEM. Three adaptive multilooking configurations were applied to generate backscatter time series for each polarization with resolutions of 20 m, 40 m, and 60 m, respectively. 

Interferometric processing produced 36 coherence maps per polarization, with temporal baseline ranging from 14-
days up to a maximum of 112 days. To ensure consistency with the three backscatter resolutions, coherence was estimated using the following number of looks: 
\begin{itemize}
    \item 21 looks (3 range × 7 azimuth),
    \item 60 looks (4 range × 10 azimuth), and
    \item 119 looks (7 range × 17 azimuth)
\end{itemize}
The resulting coherence maps were resampled to 20 m, 40 m, and 60 m resolutions, respectively, using bilinear interpolation, to perfectly align with the backscatter grid.

Using multiple resolutions enables testing the ability of deep learning models to provide reliable high-resolution tree-height estimates while coping with speckle and interferometric coherence biases. 

In addition to pairwise coherences, the median temporal coherence was computed for each time lag and polarization, and the temporal median backscatter maps were also derived. Geometry layers obtained during the SAR processing, including the local incidence angle and layover–shadow masks, were also generated and used as ancillary information in the analysis. All intensity and coherence layers were geocoded to WGS84 (EPSG:4326) reference system and re-sampled to a regular grid with 20 m pixel spacing using bilinear interpolation, while nearest neighbor interpolation was used for resampling mask layers.

\begin{figure*}[t]
    \centering    \includegraphics[width=0.8\linewidth]{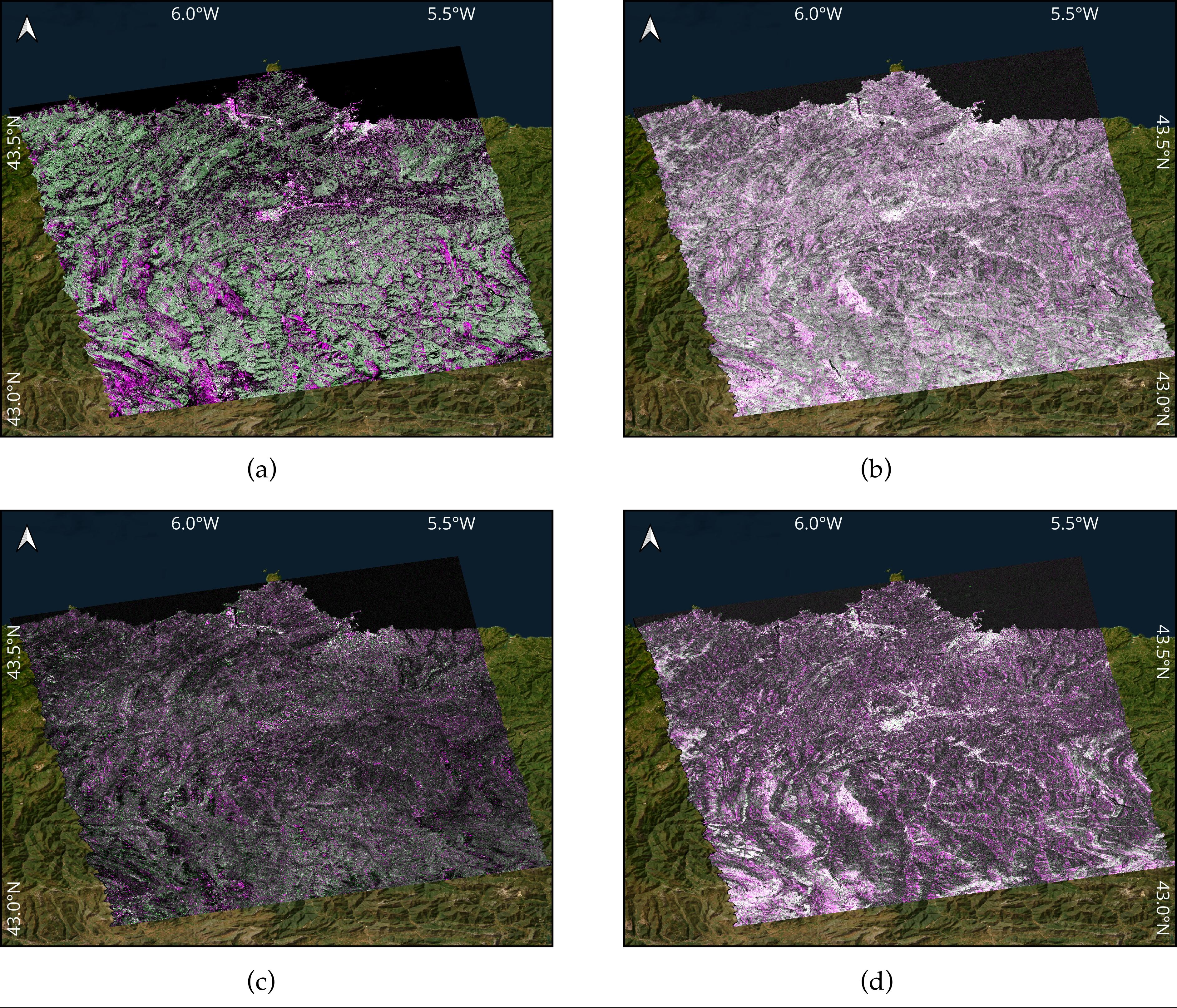}
    \caption{ RGB composites of the 20 m resolution InSAR dataset and model derived layers used in the analysis (R: HH, G: HV, B: HH). (a) Median backscatter maps in dB scale, with co- and cross-polarization intensities ranging from –14 to –4 dB and –20 to –10 dB, respectively. (b) Median 14-day coherence maps, with values between 0.2 and 0.8. (c) Temporal decay parameters, $\tau$, spanning 1–45 days. (d) Long-term coherence, $\rho_\infty$, ranging 0.2–0.7. 
}
    \label{fig:sar_data}
\end{figure*}

\begin{figure*}[t]
    \centering
    \includegraphics[width=0.9\linewidth]{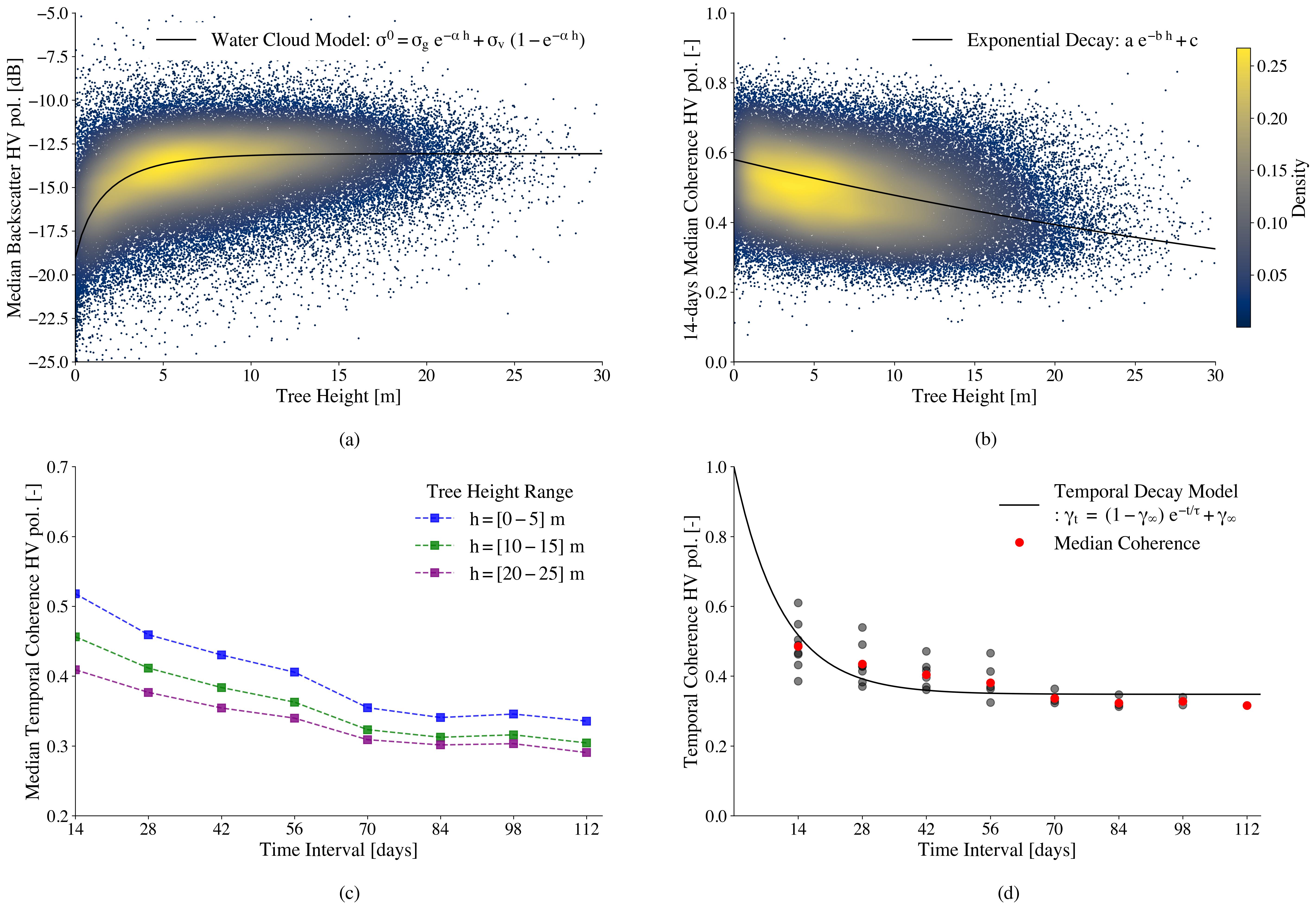}
    \caption{Top row: Sensitivity of median HV backscatter (left) and 14-day HV temporal coherence (right) to reference tree height. The backscatter is fitted using the Water Cloud Model (WCM), while the coherence is fitted with an exponential decay function, highlighting its dependence on forest structure. Bottom row: Sensitivity of ALOS-2 HV temporal coherence to temporal baseline. The left panel shows the median HV coherence as a function of repeat interval for different tree-height intervals, while the right panel presents a pixel-level example of temporal coherence decay, showing individual HV interferometric pairs (gray dots), the median stack (red dots), and the fitted exponential decay model (solid line) following \cite{sica2019repeat}.  Only pixels classified as vegetated were included in the analysis.
    }\label{fig:features_sentivity}
\end{figure*}
\subsubsection{Model-derived features}
\label{subsec:model_der_features}

Given the interferometric geometry ($k_z^{max} = 0.02~\mathrm{rad ~ m^{-1}}$), the contribution of volume decorrelation to the measured coherence is negligible \citep{kugler2015forest}. Therefore, the dependence of coherence on tree height is driven primarily by temporal decorrelation.

As reported in the literature \citep{lavalle2011temporal}, the impact of temporal decorrelation is expected to vary with tree height, as taller canopies are more generally affected by wind-induced motion. 

To capture this behavior, the SAR dataset includes parameters estimated from an exponential decay model of temporal coherence as a function of repeat interval \citep{sica2019repeat}
\begin{equation}
   \gamma_t(t) = (1- \rho_\infty) e^{-t/\tau} + \rho_\infty
\end{equation}
where $\gamma_t(t)$ is the temporal coherence, and $\rho_{\infty}$ and $\tau$ are the long-term coherence and the coherence rate of decay, respectively. The resulting maps, $\rho_{\infty}$ and $\tau$, are expected to capture the temporal and spatio-temporal variability of the interferometric coherence at the pixel level, thus to be sensitive to forest height. Model fitting was performed using median-aggregated coherence estimates at the  different spatial resolutions (20 m, 40 m, and 60 m) through a trust-region reflective regression algorithm. 

The InSAR features, including backscatter intensities, interferometric coherence layers, and model-derived parameters, are illustrated in Fig.~\ref{fig:sar_data}. Figure~\ref{fig:features_sentivity} presents the sensitivity of the 20 m resolution median HV backscatter to reference tree height (Figure~\ref{fig:features_sentivity},a), the sensitivity of the 14-day temporal median coherence to tree height (Figure~\ref{fig:features_sentivity},b), and the temporal decay of coherence over increasing baselines (Figure~\ref{fig:features_sentivity}-c,d). Figure~\ref{fig:features_sentivity}-d provides an example of a pixel-level fit of the temporal coherence decay model.

\subsection{Reference data}
\label{subsec:ref_data}
Reference tree height data are provided by airborne LiDAR-derived forest height measurements. These data come from the Normalized Digital Surface Model of Vegetation Classes of Spain (MDSnV), generated from the second national LiDAR coverage acquired between 2015 and 2021, with an original spatial resolution of 2.5 m \citep{MDSnV2021}. The MDSnV was obtained by interpolating vegetation height above ground from LiDAR returns classified into low, medium, and tall vegetation classes (C3, C4, and C5, respectively) as part of the Spanish National Plan for Aerial Orthophotography (PNOA). For the study site, LiDAR data were collected in 2020, ensuring temporal consistency with SAR acquisitions from the same year. 

The mean tree height at the study site was 7.80 m.  
The reference data were aggregated into 20 m, 40 m, and 60 m mapping units to match the spatial resolutions of the input features.

\section{Methods}\label{sec:methods}

\subsection{General approach}

This section presents the proposed methodology, describing the adopted deep learning framework, the training and validation strategies, and the final inference and reliability assessment steps.

The input feature set comprises interferometric and ancillary layers derived from ALOS-2 data, which were used either individually or in combination to predict forest height. Specifically, the considered features include:

\begin{itemize}
    \item Median backscattering coefficients in HH and HV polarizations
    \item Median HV and HH temporal coherence time series, with temporal baselines ranging from 14 to 112 days;
    \item Model-derived parameters $\tau$ and $\rho_\infty$ 
    \item Local incidence angle;
    \item Copernicus DEM at 30 m spatial resolution.
\end{itemize}

A family of U-Net–based architectures was adopted as the backbone for the proposed SAR-time-series based forest height mapping framework. The tested models include the original U-Net model \citep{ronneberger2015unet}, Nested U-Net \citep{zhou2018unet++}, and SeU-Net \citep{ge2022improved}, which incorporates squeeze-excitation blocks \citep{hu2020seunet}. These architectures were selected for their capability to effectively model spatially structured patterns from limited reference data, a common constraint in remote sensing forest inventory mapping tasks. 

To evaluate the performance of the proposed models, we conducted experiments using different combinations of input features and spatial resolutions.
Each configuration was initially trained and evaluated at 20~m resolution, while additional experiments at 40~m and 60~m were performed for selected representative cases. For these cases, conventional machine learning algorithms, namely Random Forest (RF) and k-Nearest Neighbors (kNN), were also implemented to provide a baseline comparison against the proposed deep learning models.

\begin{figure}[h!]
    \centering
    \includegraphics[width=1\linewidth]{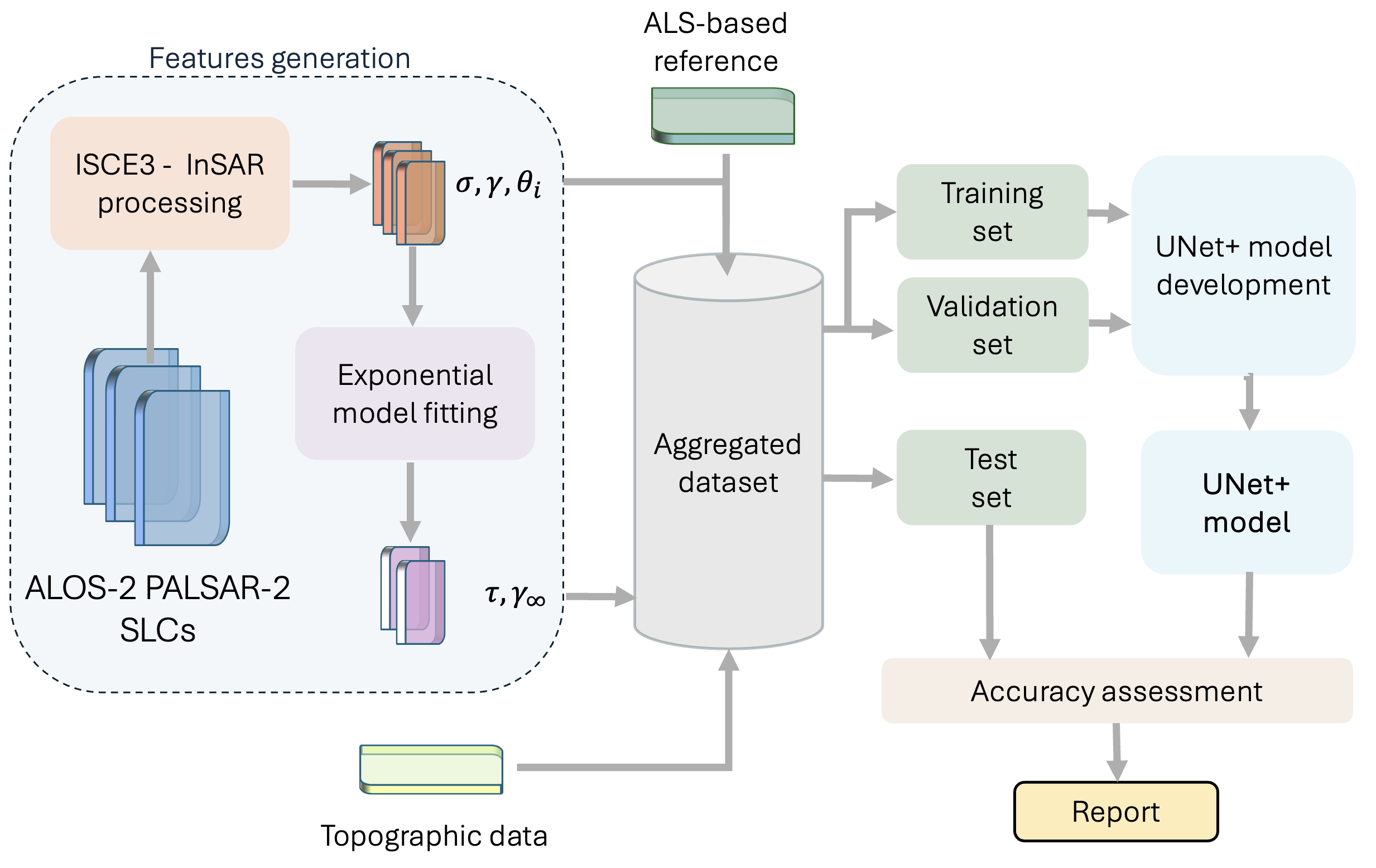}
    \caption{General workflow of the proposed hybrid deep learning approach for forest height mapping from multi-temporal PALSAR-2 interferometric and backscatter features.}
    \label{fig:studylogic}
\end{figure}

The overall workflow of the study is shown in Fig. \ref{fig:studylogic}.

\subsection{UNet modeling}

The U-Net architecture is essentially a convolutional encoder-decoder network originally developed for biomedical image segmentation \citep{ronneberger2015unet}. It has since been widely applied to EO-based semantic segmentation and regression tasks \citep[e.g.,][]{Scepanovic2021,ge2022improved, ge_transfer2023, illarionova2022,bolyn2022}. Its encoder-decoder design, reinforced by skip connections, enables the preservation of fine spatial details and the reconstruction of high-resolution output maps, essential for predicting forest attributes such as forest height, growing stock volume and above ground biomass density.

\begin{figure*}[h!]
    \centering\includegraphics[width=1\linewidth]{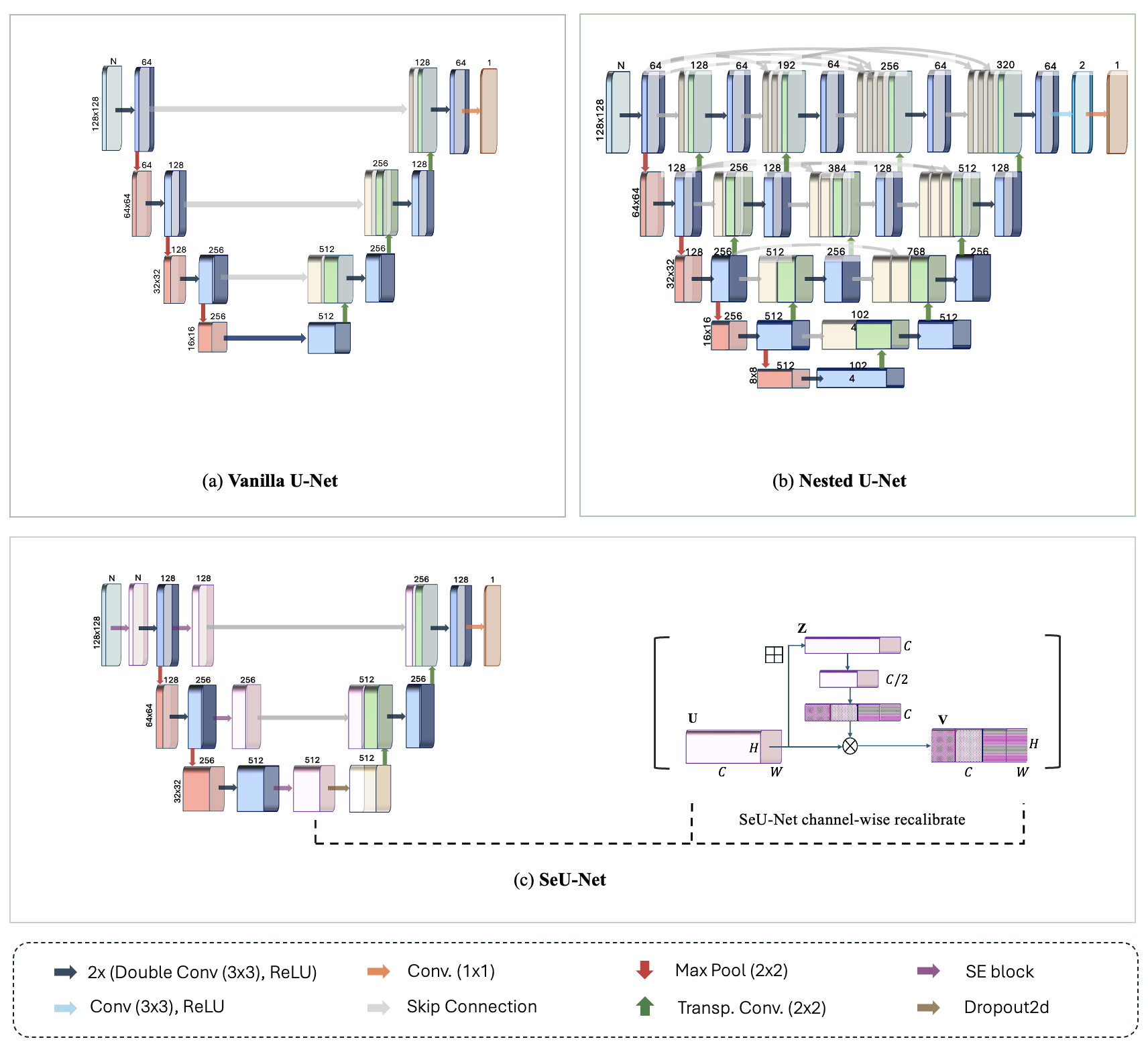}
    \caption{Structure of studied U-Net models, where where the Vanilla U-Net (a) represents the baseline design, SeU-Net (c) incorporates squeeze-and-excitation blocks for channel-wise attention, and Nested U-Net (b) introduces dense skip connections}
    \label{fig:allunetmodels}
\end{figure*}

The Vanilla U-Net consists of a symmetrical encoder--decoder structure composed of double convolutional blocks, each including a 2-D convolution, batch normalization, and ReLU activation. This design allows the model to extract deep hierarchical features while maintaining spatial correspondence between input and output layers, making it well suited for pixel-level regression and classification tasks (Figure~\ref{fig:allunetmodels}a).

Building on this foundation, several U-Net family variants were explored to enhance the representation of forest structural heterogeneity in spatially explicit EO data:

\begin{itemize}
    
    \item Nested U-Net introduces densely connected skip pathways that improve multi-scale feature fusion and gradient propagation, enhancing robustness in heterogeneous forest environments where spectral and structural complexity may lead to ambiguous signals (Fig.\ref{fig:allunetmodels},b).
    \item SeU-Net extends the U-Net design by incorporating squeeze-and-excitation blocks, which enable the network to adaptively recalibrate channel-wise feature responses. This mechanism is particularly beneficial for multi-source input images, where the relative importance of different SAR and ancillary inputs can vary across forest types and ecological and seasonal conditions. The adoption of SE blocks has been shown to yield superior accuracy in boreal forest height mapping using Sentinel-1 and Sentinel-2 time series \citep{ge2022improved} and in other studies employing multi-feature datasets \citep{Gazzea2023} (Fig. \ref{fig:allunetmodels},c).

\end{itemize}

\subsection{Experimental design and setup}

This section outlines the organization of the experiments performed with the preprocessed SAR feature stack and reference data set described in Sections \ref{subsec:sar_aux_data}, \ref{subsec:model_der_features} and \ref{subsec:ref_data}. 

The whole study area was divided into image patches of size 128 × 128 px, with pixel size of 20~$m$, thus size of each patch was 2.56 x 2.56 $km^{2}$. To train, validate, and test the studied models, the input dataset was divided into three subsets: 60\% for training, 20\% for validation, and 20\% for testing, using random cropping. Before training, all input bands were normalized using Z-score based normalization using band-wise mean and standard deviation.

The models were trained using a masked mean squared error (masked-MSE) loss function, which computes the prediction error only over valid reference pixels. Invalid or missing-data regions were excluded, such as radar layover/shadow areas. The loss function is defined as

\begin{equation}
\mathcal{L}(\boldsymbol{\theta}) =
\frac{\sum_{i=1}^n  \big(f_{\boldsymbol{\theta}}(x_i) - y \big)^2}{n}
\end{equation}
where \(x_i\) represents the input features for valid pixel \(i\), \(y_i\) is the 
corresponding reference tree height, \(f_{\boldsymbol{\theta}}(x_i)\) is the 
model prediction parameterized by the network weights \(\boldsymbol{\theta}\).

\begin{figure}[h!]
    \centering
\includegraphics[width=1\linewidth]{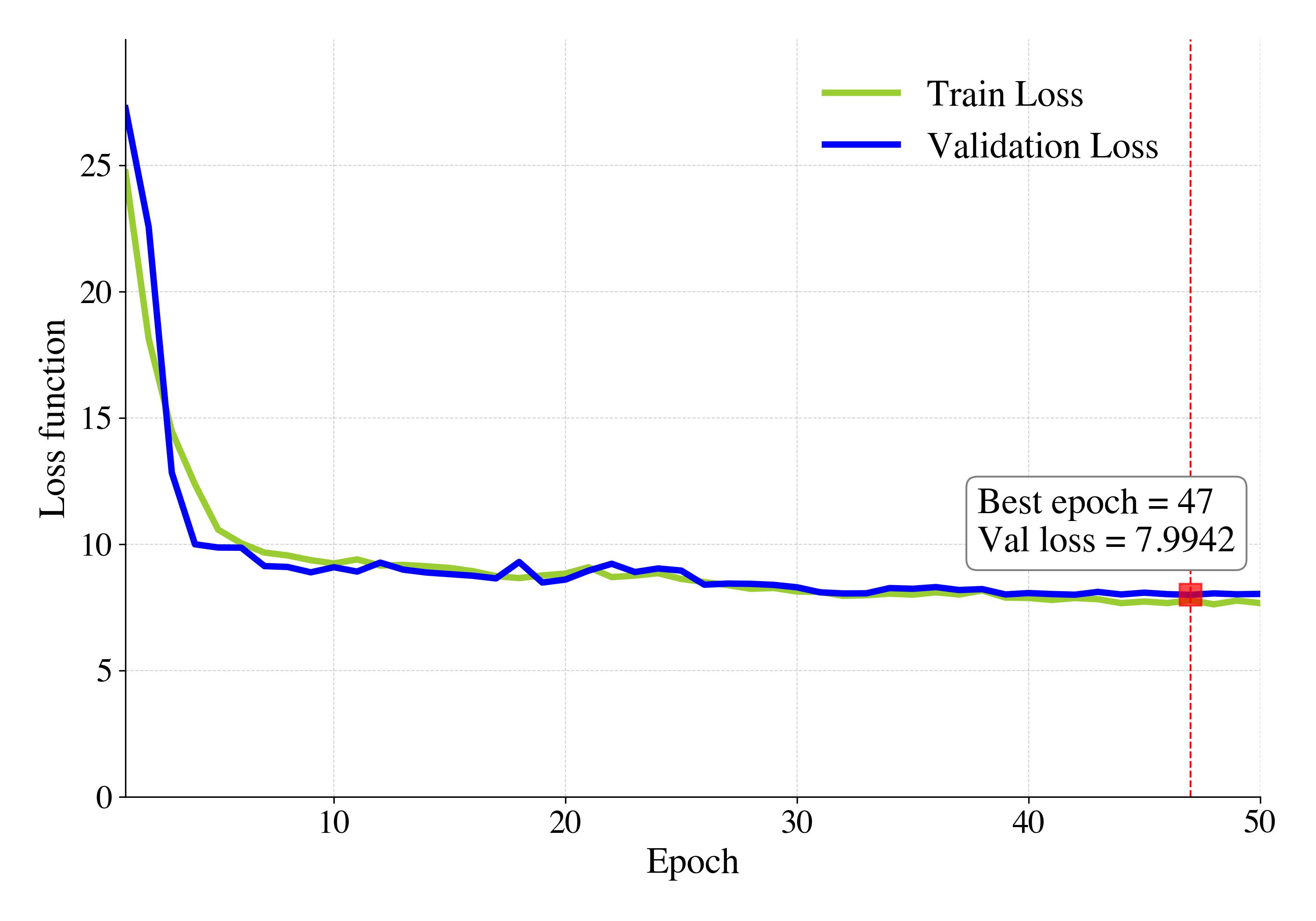}
    \caption{Accuracy curves during Nested UNet model training and development using all available input features.}
    \label{fig:trainvalloss}
\end{figure}

Experiments were run on a Google Colab platform with Tesla T4 GPUs. The proposed model was built by PyTorch 2.8 with CUDA 12.6 and Python 3.12 . 
During the training, Adam \citep{kingma2017adammethodstochasticoptimization} was chosen as the optimizer and OneCycleLR as the learning rate scheduler, the initial learning rate was set to 10$^{-3}$, and patience parameter was set to 5. We trained the model for 50 epochs, with batch size of 16, the weight decay factor was set to 10$^{-4}$. The checkpoints were saved and updated according to the validation loss, with the best checkpoint used in the testing stage. An example of training and validation loss functions is shown in Figure~\ref{fig:trainvalloss}.

\subsection{Accuracy metrics}
To evaluate model performance, we employed the Mean Error (ME), Root Mean Square Error (RMSE), and the coefficient of determination ($R^{2}$). Let $y_i$ denote the reference values, $\hat{y}_i$ the predicted values, and $n$ the total number of samples.
Mean Error represents the average deviation between predicted and reference values:

\begin{equation}
\text{ME} = \frac{1}{n} \sum_{i=1}^{n} (\hat{y}_i -y_i)
\end{equation}

RMSE measures the overall magnitude of prediction errors, with larger errors contributing more strongly to the metric due to the squared residuals:
\begin{equation}
\text{RMSE} = \sqrt{ \frac{1}{n} \sum_{i=1}^{n} (\hat{y}_i -y_i)^2 }
\end{equation}

Coefficient of Determination $R^2$ quantifies the proportion of variance in the reference values explained by the predictions:
\begin{equation}
R^2 = 1 - \frac{\sum_{i=1}^{n} (\hat{y}_i -y_i)^2}{\sum_{i=1}^{n} (y_i - \bar{y})^2}
\end{equation}
where $\bar{y}$ is the mean of the reference values.

\section{Results and Discussion}\label{sec:results}

\begin{figure*}[h]
    \centering
    \includegraphics[width=1\linewidth]{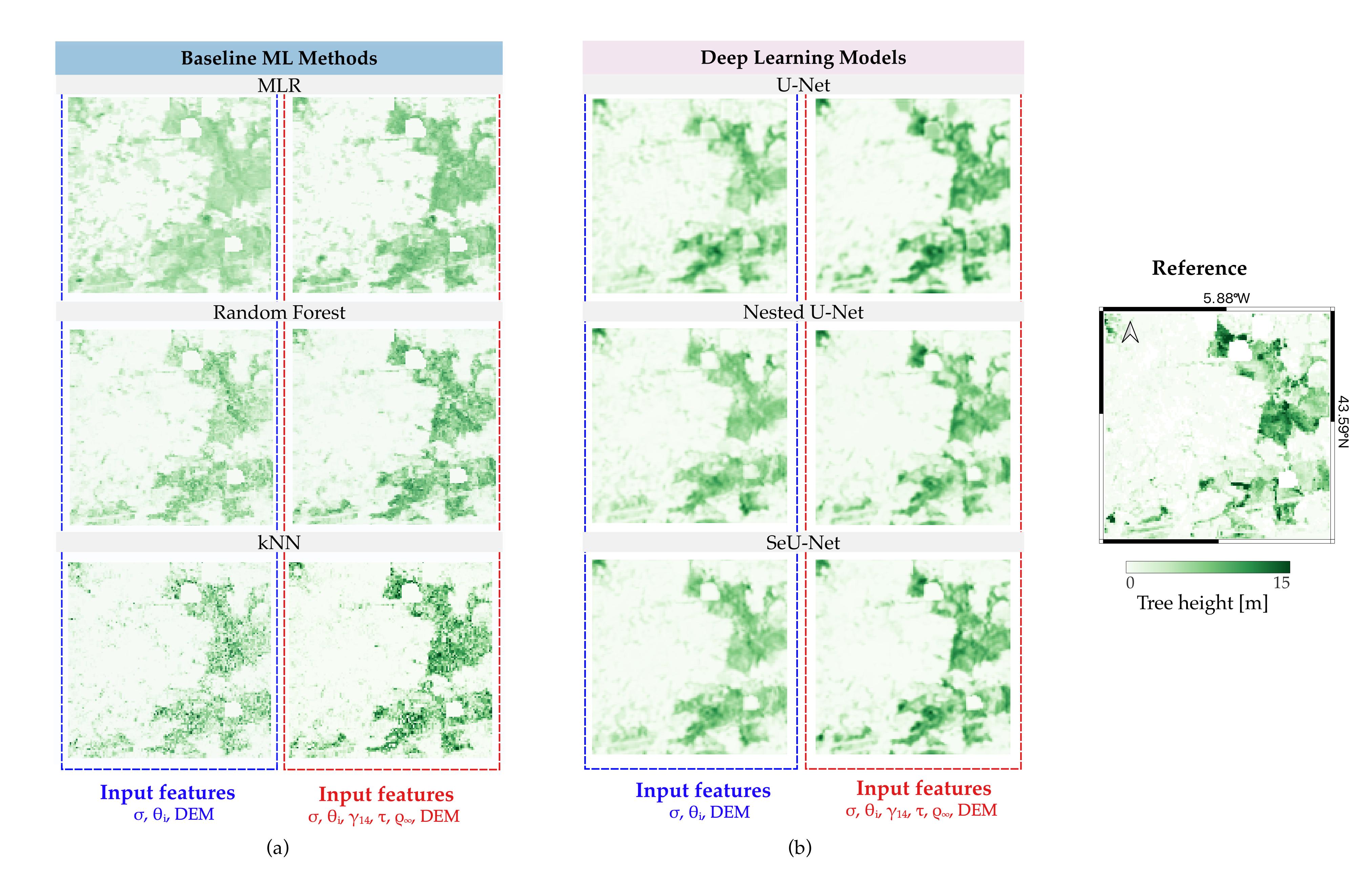}
    \caption{Example of tree height estimates over a forest patch obtained with baseline machine learning (a) and deep learning (b) models using two input feature configurations at 20 m resolution. }
    \label{fig:patches}
\end{figure*}

\begin{figure*}[t]
    \centering
    \includegraphics[width=0.9\linewidth]{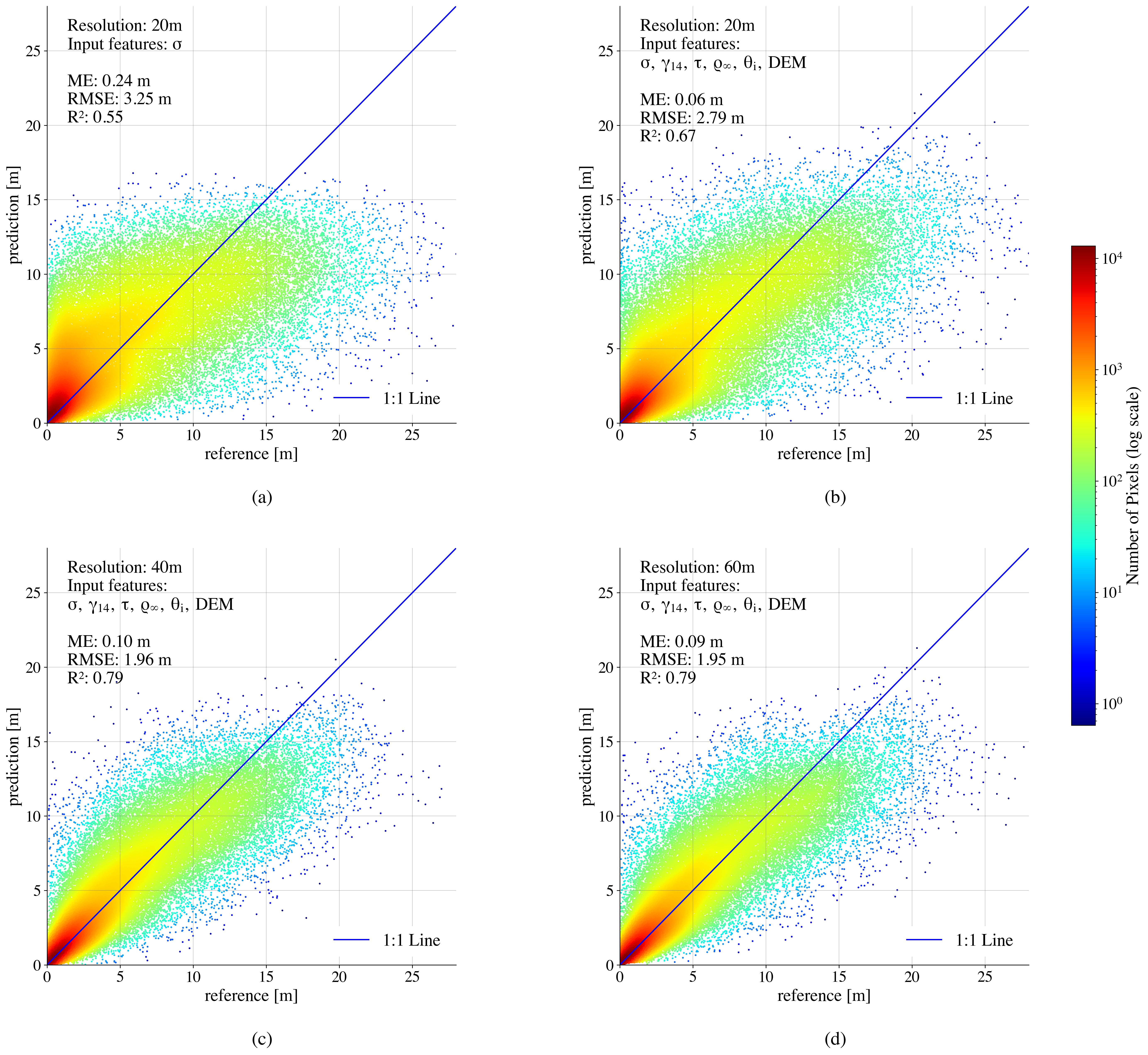}
    \caption{Scatterplots showing the effect of input feature configuration and spatial resolution on SeU-Net tree-height predictions using HH and HV polarizations. (a) 20 m resolution using backscatter ($\sigma$) only;
(b) 20 m resolution using all features ($\sigma$, $\gamma_{14}$, $\tau$, $\theta_i$, DEM);
(c) 40 m resolution using all features;
(d) 60 m resolution using all features.
The 1:1 line is shown in blue for reference.}
    \label{fig:perfomance_204060m_scatterplots}
\end{figure*}

\begin{figure}[t]
    \centering    \includegraphics[width=1\linewidth]{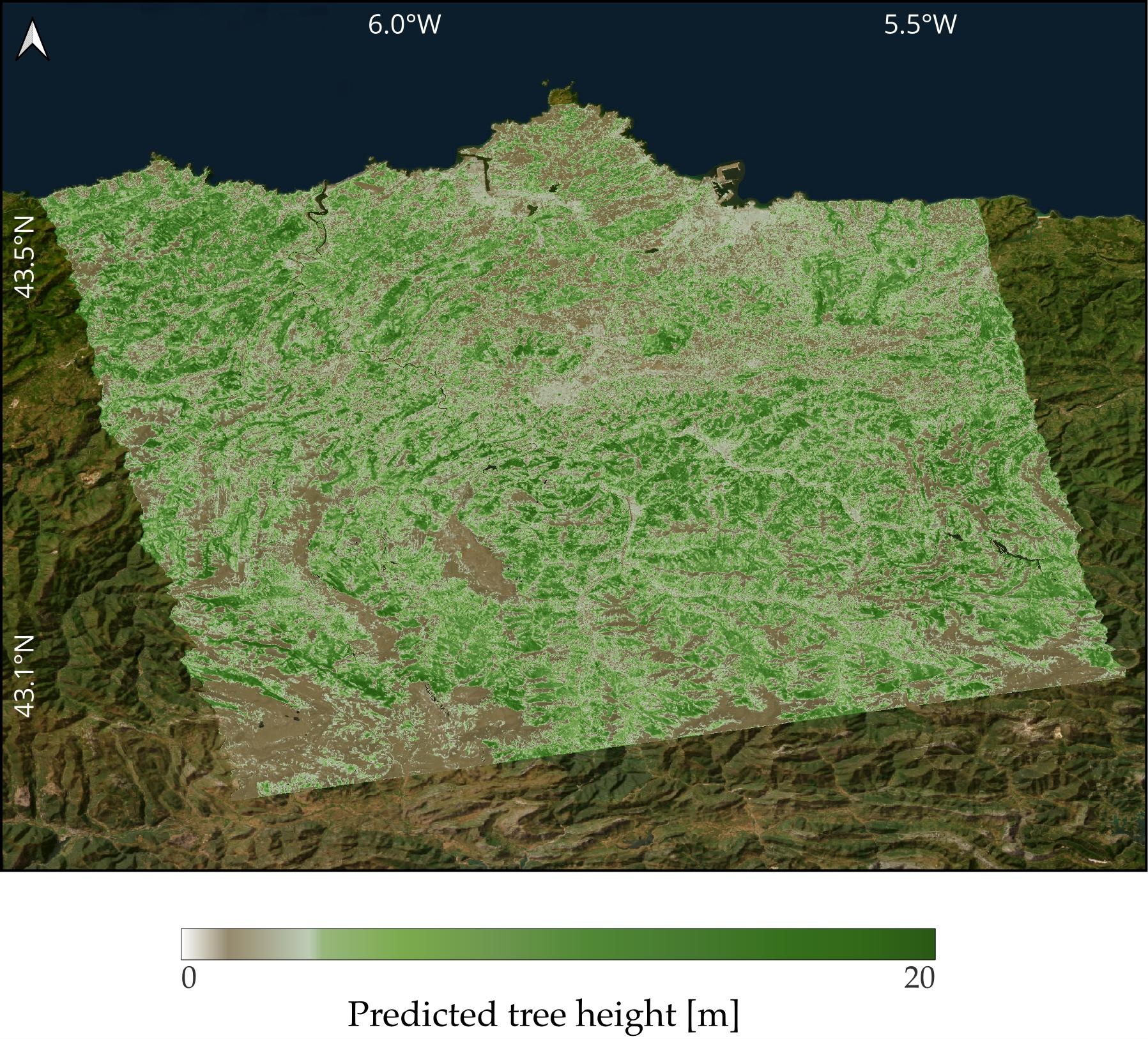}
    \caption{Tree height predictions from the SeU-Net model at 20 m resolution using median backscatter, 14-day coherence, and model-derived layers ($\tau$, $\rho_\infty$) at co- and cross-polarizations.}
    \label{fig:predicted_tree_height}
\end{figure}

This section presents the results obtained from the proposed methodology, including the performance of the deep learning architectures across different input configurations and spatial resolutions. The tree height estimation results for the 20 m resolution experiments are reported in Table~\ref{tab:tab1_20}, while Table~\ref{tab:tab2_204060} summarizes the outcomes obtained at 20 m, 40 m, and 60 m resolutions for selected input feature configurations. Examples of forest patch predictions are shown in Fig.~\ref{fig:patches}, and selected scatterplots illustrating model performance  are presented in Fig.~\ref{fig:perfomance_204060m_scatterplots}. A map of the retrieved tree height obtained using the SeU-Net model at 20 m resolution and all available input features is shown in Fig.~\ref{fig:predicted_tree_height}.

\begin{table*}[ht!]
    \centering
    \resizebox{0.85\textwidth}{!}{
    \begin{tabular}{c|c|ccc|ccc|ccc}
             \multicolumn{11}{c}{\textbf{Performance Metrics}} \\
 \hline
        Input                                                           & Model                        & \multicolumn{3}{c|}{ME [m]}    & \multicolumn{3}{c|}{RMSE [m]}   &    \multicolumn{3}{c}{R2 [-]} \\
                                                                        &                              & HH     & HV     & HH,HV        & HH     & HV     & HH,HV         & HH     & HV     & HH,HV    \\
        \hline
        \multicolumn{11}{c}{\textbf{single-features}} \\
\hline

  \multirow{3}{*}{$\sigma$}                                             & UNet                         &0.25    & 1.36   &0.20          &3.77    & 3.82   &3.49           &0.39    &0.38    &0.47  \\
                                                                        & Nested UNet                       &0.18    & 0.21   &0.05          &3.48    &3.29    &\textbf{3.17}           &0.48    &0.53    &0.57  \\
                                                                        & SeU-Net                       &0.12    & 0.23   & 0.24         &3.54    &3.25    &3.25           &0.46    &0.55    &0.55  \\

    \hline                         
 \multirow{3}{*}{$\gamma_{14}$}                                         & UNet                         & 0.23   &0.18    &-0.04         &3.29    &4.26   &3.90           &0.54    &0.22      &0.35  \\
                                                                        & Nested UNet                       & 0.15   &0.00    &0.26          &3.97    &4.23   &\textbf{3.81}           &0.33    &0.23      &0.38  \\
                                                                        & SeU-Net                       & 0.08   &0.29    &0.27          &3.97    &4.19   &3.85           &0.33    &0.25      &0.37  \\
        \hline 
 \multirow{3}{*}{$\tau, \rho_\infty$}                                   & UNet                         & -0.02  & 0.10   & 0.17         &4.02    &4.14   &4.01           &0.31    &0.29      &0.31  \\
                                                                        & Nested UNet                       & 0.11   & -0.12  & 0.04         &3.91    &4.05   &\textbf{3.80}           &0.35    &0.30      &0.38  \\
                                                                        & SeU-Net                       & -0.01  & 0.08   & 0.05         &3.91    &4.05   &3.81           &0.35    &0.30      &0.39  \\
        \hline                            
$\gamma_{14}, \gamma_{28}, \gamma_{42},\gamma_{56}$,                    & UNet                         &0.20    &-0.02   &0.08          &3.89    &3.98   &3.79           &0.35    &0.32      &0.39  \\
$\gamma_{70}, \gamma_{84}, \gamma_{90}, \gamma_{112}$                   & Nested UNet                       &0.21    &-0.15   &0.13          &3.75    &3.92   &\textbf{3.62}           &0.40    &0.34      &0.44  \\
                                                                        & SeU-Net                       &0.08    &0.19    &0.18          &3.70    &3.91   &3.66           &0.41    &0.34      &0.43  \\
            \hline

            \multicolumn{11}{c}{\textbf{combined-features}} \\
\hline

  \multirow{3}{*}{$\sigma$,$\gamma_{14}$}                               & UNet                         & 0.24   & 0.13   &0.16          &3.29    &3.20   &3.19           &0.54    &0.56      &0.56  \\
                                                                        & Nested UNet                       & 0.08   & -0.01  &0.16          &3.13    &3.00   &3.00           &0.58    &0.61      &0.61  \\
                                                                        & SeU-Net                       & -0.15  & 0.07   &0.00          &3.15    &3.09   &\textbf{2.94}           &0.57    &0.59      &0.63  \\

    \hline                         
 \multirow{3}{*}{$\sigma$,$\tau, \rho_\infty$}                          & UNet                         & 0.12   &0.46    &0.28          &3.29    &3.34   &3.11           &0.54    &0.52      &0.59  \\
                                                                        & Nested UNet                       & 0.05   &0.00    &0.00          &3.13    &3.04   &\textbf{2.94}           &0.58    &0.60      &0.63  \\
                                                                        & SeU-Net                       & 0.00   &0.04    &0.04          &3.12    &2.96   &2.96           &0.58    &0.62      &0.62  \\
        \hline 
 \multirow{3}{*}{$\sigma$,$\gamma_{14}$,$\tau, \rho_\infty$}            & UNet                         & -0.02  &-0.10   &-0.03         & 3.30   &3.16   &3.03           & 0.53   &0.57      &0.61  \\
                                                                        & Nested UNet                       & 0.26   & 0.17   &0.10          & 3.13   &2.94   &\textbf{2.89}           & 0.58   & 0.63     &0.64  \\
                                                                        & SeU-Net                       & 0.05   &-0.04   &0.06          & 3.04   &2.96   &2.92           & 0.60   & 0.62     &0.63  \\
        \hline 
$\sigma$,                           
$\gamma_{14}, \gamma_{28}, \gamma_{42},\gamma_{56}$,                    & UNet                         &0.16    &0.06    &0.00          &3.22    &3.07   &3.03           &0.56    &0.60      &0.61  \\
$\gamma_{70}, \gamma_{84}, \gamma_{90}, \gamma_{112}$                   & Nested UNet                       &-0.02   &-0.09   &0.08          &3.06    &2.94   &\textbf{2.90}           &0.60    &0.63      &0.64  \\
                                                                        & SeU-Net                       &0.10    &0.16    &-0.05         &3.00    &2.93   &2.92           &0.61    &0.63      &0.63  \\
            \hline
        
            \multicolumn{11}{c}{\textbf{combined-features + geometry layers}} \\
\hline                         
 \multirow{3}{*}{$\sigma ,  \theta_i$}                                  & UNet                         &0.05    &0.24    &0.25          &3.60    &3.31    &3.25           &0.44    &0.53    &0.55  \\
                                                                        & Nested UNet                       &-0.09   &-0.08   &0.09          &3.15    &3.02    &\textbf{2.95}           &0.57    &0.61    &0.63  \\
                                                                        & SeU-Net                       &0.14    &0.11    &0.08          &3.22    &3.11    &3.02           &0.56    &0.59    &0.61  \\
        \hline 
 \multirow{3}{*}{$\sigma ,  \theta_i$, dem}                             & UNet                         &0.01    &-0.12   &0.15          &3.28    &3.15    &3.13           &0.54    &0.58    &0.58     \\
                                                                        & Nested UNet                       &-0.06   &0.11    &0.11          &3.15    &2.98    &\textbf{2.93}           &0.54    &0.62    &0.63     \\
                                                                        & SeU-Net                       &0.00    &0.01    &0.18          &3.11    &2.98    &2.94           &0.58    &0.62    &0.63      \\
        \hline                            
   \multirow{3}{*}{$\sigma , \theta_i$ ,dem, $\gamma_{14}$}             & UNet                         &-0.08   &0.13    &0.00          &3.13    &3.01    &2.84           &0.58    &0.61    &0.66  \\
                                                                        & Nested UNet                       &0.24    &0.27    &0.01          &2.86    &2.87    &\textbf{2.78}           &0.62    &0.65    &0.67    \\
                                                                        & SeU-Net                       &0.03    &0.05    &0.03          &3.00    &2.88    &2.82           &0.61    &0.64    &0.66     \\
            \hline
\multirow{3}{*}{$\sigma , \theta_i$ ,dem, $\gamma_{14}, \tau, \rho_\infty$}&UNet                       &0.05    &0.10    &-0.09         &2.98    &2.90    &2.93           &0.62    &0.64    &0.63    \\
                                                                        & Nested UNet                       &0.05    &0.01    &0.04          &2.92    &2.83    &\textbf{2.77}           &0.63    &0.66    &0.67    \\
                                                                        & SeU-Net                       &0.05    &0.01    &0.06          &2.92    &2.79    &2.79           &0.64    &0.67    &0.67     \\
    \hline
\multirow{3}{*}{$\sigma , \theta_i , \tau, \rho_\infty$}                & UNet                         &0.09    &0.14    &0.04          &3.36    &3.12    &3.10           &0.52    &0.58    &0.57    \\
                                                                        & Nested UNet                       &-0.02   &0.00    &0.07          &2.95    &2.88    &2.83           &0.63    &0.64    &0.66    \\
                                                                        & SeU-Net                       &0.12    &0.06    &0.00          &2.98    &2.86    &\textbf{2.82}           &0.62    &0.65    &0.66    \\
        \hline
$\sigma , \theta_i$, dem                                                & UNet                         &0.21    &0.23    &0.14          &3.16    &3.09    &2.83           &0.57    &0.59    &0.66    \\
$\gamma_{14}, \gamma_{28}, \gamma_{42},\gamma_{56}$,                    & Nested UNet                       &0.21    &-0.09   &0.04          &2.93    &2.83    &2.79           &0.63    &0.66    &0.67    \\
$\gamma_{70}, \gamma_{84}, \gamma_{90}, \gamma_{112}$                   & SeU-Net                       &0.16    &0.00    &0.17          &2.88    &2.82    &\textbf{2.78}           &0.64    &0.66    &0.67    \\
        \hline
    \end{tabular}
    }
    \caption{Prediction performance of all evaluated deep learning models across different SAR feature modalities at 20 m spatial resolution. Best achieved performance for each input modality is highlight in bold}
    \label{tab:tab1_20}
\end{table*}

\begin{table*}[ht]
    \centering
    \resizebox{0.75\textwidth}{!}{
    \begin{tabular}{c|c|ccc|ccc|ccc}
        \hline
             \multicolumn{11}{c}{\textbf{Performance Metrics}} \\
 \hline
        Input                                                           & Model                        & \multicolumn{3}{c|}{ME [m]}    & \multicolumn{3}{c|}{RMSE [m]}   &    \multicolumn{3}{c}{R2 [-]} \\
                                                                        &                              & 20m     & 40m   & 60m          & 20m     & 40m   & 60m           & 20m     & 40m   & 60m    \\
       \hline
 \multirow{6}{*}{$\sigma ,  \theta_i$, dem}                             & RF                           &0.12     &0.08   &0.08          &3.90    &2.40    &2.40           &0.29     &0.66    &0.66\\
                                                                        & MLR                          &0.20     &0.35   &0.35          &3.32    &3.32    &3.32           &0.34     &0.34    &0.34\\
                                                                        & kNN                          &0.07     &0.05   &0.05          &3.62    &2.60    &2.60           &0.59     &0.58    &0.59\\
                                                                        & UNet                         &0.15     &0.13   & 0.06         &3.13    &2.36    & 2.38          &0.58     &0.69    &0.69\\
                                                                        & Nested UNet                  &0.11     &0.07   & 0.10         &\textbf{2.93}    &\textbf{2.19}    & \textbf{2.19}          &0.63     &0.73    &0.74\\
                                                                        & SeU-Net                       &0.18     &0.21   & 0.07         &2.94    &2.24    & 2.23          &0.63     &0.72    &0.73\\
        
    \hline
    \multirow{6}{*}{$\sigma,\gamma_{14},  \theta_i$, dem}               & RF                           &0.08     &0.04   &0.04          &3.10    &2.17    &2.16            &0.55    &0.71   &0.72\\      
                                                                        & MLR                          &0.04     &0.06   &0.05          &3.50    &2.76    &2.74            &0.43    &0.54   &0.55\\
                                                                        & kNN                          &0.03     &0.02   &0.01          &3.36    &2.38    &2.38            &0.47    &0.66   &0.66\\
                                                                        & UNet                         &0.00     &0.02   &0.01          &2.84    &2.12    &2.06            &0.66    &0.75   &0.76\\
                                                                        & Nested UNet                  &0.01     &0.06   &-0.01         &\textbf{2.78}    &\textbf{2.03}    &\textbf{2.03}            &0.67    &0.77   &0.77\\
                                                                        & SeU-Net                       &0.03     &0.10   &0.09          &2.82    &2.07    &2.07            &0.66    &0.76   &0.76\\

    \hline
\multirow{6}{*}{$\sigma , \theta_i$ ,dem, $\gamma_{14}, \tau, \rho_\infty$}& RF                        &0.07     &0.06   &0.07          &3.17    &2.12   &2.12            &0.54     &0.72    &0.73\\
                                                                        & MLR                          &0.04     &0.09   &0.07          &3.48    &2.74   &2.73            &0.44     &0.55    &0.56\\
                                                                        & kNN                          &0.01     &0.05   &0.05          &3.46    &2.37   &2.36            &0.45     &0.66    &0.67\\
                                                                        & UNet                         &-0.09    &-0.03   &-0.03        &2.93    &2.14   &2.14            &0.63     &0.75    &0.75 \\
                                                                        & Nested UNet                  &0.04     &0.03    &0.02         &\textbf{2.77}    &1.97   &1.98            &0.67     &0.78    &0.78 \\
                                                                        & SeU-Net                       &0.06     &0.10    &0.09         &2.79    &\textbf{1.96}   &\textbf{1.95}            &0.67     &0.78    &0.79 \\
    \hline
                                                                        &RF                            &0.12     &0.07    &0.07         &3.06    &2.12   &2.11            &0.56     &0.73   &0.73 \\
    $\sigma , \theta_i$, dem                                            & MLR                          &0.06     &0.07    &0.07         &3.48    &2.75   &2.73            &0.44     &0.55   &0.55 \\
    $\gamma_{14}, \gamma_{28}, \gamma_{42},\gamma_{56}$,                & kNN                          &0.07     &0.06    &0.06         &3.44    &2.47   &2.43            &0.45     &0.63   &0.64 \\
    $\gamma_{70}, \gamma_{84}, \gamma_{90}, \gamma_{112}$               & UNet                         &0.14     &0.07    &0.12         &2.83    &2.02   &2.04            &0.66     &0.77   &0.77 \\
                                                                        & Nested UNet                  &0.04     &0.10    &0.02         &2.79    &2.03   &\textbf{1.99}            &0.67     &0.77   &0.78 \\
                                                                        & SeU-Net                       &0.17     &0.14    &0.11         &\textbf{2.78}    &\textbf{1.98}   &2.00            &0.67     &0.77   &0.79 \\
\hline
    \end{tabular}
    }

    \caption{Prediction performance of all evaluated deep learning models at 20m, 40m and 60m resolution spatial resolutions using dual pol (HH, HV) input features.  Best achieved performance for each input modality is highlight in bold.}
    \label{tab:tab2_204060}
\end{table*}

\subsection{General prediction performance observations}

Overall, examined deep learning models demonstrated consistent capability to predict the target variable from multi-temporal PALSAR-2 datasets with high accuracy. The baseline U-Net provided a consistent reference for comparison, while two modified variants, i.e., Nested U-Net, featuring nested skip connections for improved multiscale feature fusion, and SeU-Net enhanced with Squeeze-and-Excitation attention blocks, both achieved higher predictive accuracy.
Across all feature configurations, inclusion of additional polarimetric and geometry-derived features/predictors improved prediction accuracies, confirming that compiled combination of SAR polarimetric observables and interferometric coherence features provides the richest information for learning. Root-mean-square errors (RMSE) typically ranged between 2.8--3.9~m, and coefficients of determination ($R^2$) reached up to 0.67 for the best configurations at 20 m mapping units, indicating robust model generalization. Results further improved to less than 2 m RMSE for larger mapping units (20 m--40 m) when processing was performed in coarser resolution.

\subsection{SAR backscatter data only}
When using only SAR backscatter intensities ($\sigma^{0}$), the models achieved moderate prediction performance, with RMSE between 3.1--3.9~m and $R^2$ values ranging from 0.45 to 0.55. 
Among the three architectures, Nested U-Net generally yielded the lowest errors and highest $R^2$ values, indicating that its dense skip connections effectively enhanced feature propagation and multiscale representation even from limited intensity-based information. SeU-Net achieved comparable accuracy in several cases, suggesting that the channel-wise recalibration introduced by the SE blocks also supports improved discrimination of backscatter variations. 
The baseline U-Net consistently provided lower accuracy than its two enhanced variants. Across all models, performance gains were clear when  dual-pol (HH,HV) inputs were used, confirming that the joint use of both polarizations provides a more complete scattering characterization than single-polarization inputs. 

\subsection{Intensity and coherence data}

Incorporating temporal coherence ($\gamma$) in addition to backscatter significantly improved predictive accuracy. The inclusion of $\gamma$ led to a notable reduction in RMSE (to approximately 3.0--3.3~m)  and an increase in $R^2$ to 0.60-0.63, reflecting the added sensitivity of temporal coherence to structural characteristics of the forests. The Nested U-Net architecture showed the most consistent improvement, as its nested skip connections enhanced the fusion of multiscale spatial and temporal features from amplitude and coherence inputs. The SeU-Net model had similar performance in prediction accuracy, occasionally even achieving slightly higher $R^2$ values, demonstrating that spatial attention via SE blocks complements the temporal information embedded in coherence. Overall, both architectures substantially outperformed the baseline U-Net.

While the combined use of backscatter intensity, geometric layers and insar coherence improved performance, the 14-day coherence used as a single input feature yielded comparatively low accuracy, particularly when generated with a limited number of looks. This behavior is due to the high spatial variability of the coherence measurements as well as uncompensated biases in the coherence estimation.

\subsection{Including additional variables}

Including model-derived parameters such as the InSAR coherence–based inverted parameters ($\tau$, $\rho_\infty$) yielded performance comparable to that obtained using the full temporal stack of interferometric coherence. This indicates both that the DL model is capable of directly learning the underlying physical relationships from the temporal coherence series, and that $\tau$ and $\rho_\infty$ represent meaningful descriptors of the temporal coherence sensitivity to forest structure. The main advantage of employing these two model-derived parameters lies in the reduced dimensionality of the input features, which lowers computational costs.

When adding incidence angle ($\theta_i$) and digital elevation model (DEM) layers, model performance improved further. These features provided geometric and physical context, allowing the networks to better account for topographic and incidence-angle induced variations in backscatter and coherence. The Nested U-Net achieved the lowest RMSE values (approximately 2.8--2.9~m) and highest $R^2$ (up to 0.67), closely followed by SeU-Net. Overall, this indicates that combining SAR observables with auxiliary physical variables enables the extraction of more meaningful representations. The improvement was especially evident for areas with complex terrain, where geometry-driven variations are pronounced and cannot be captured by $\sigma^{0}$ and $\gamma$ alone.

\subsection{Coarser resolution mapping}

The comparison of performances at 20 m, 40 m, and 60 m resolutions showed that all models maintained stable accuracy across scales, with improved prediction accuracy at coarser mapping units, as illustrated in Table~\ref{tab:tab2_204060} and Fig.~\ref{fig:perfomance_204060m_scatterplots}-b,c,d.

From physics of radio wave scattering based perspective, larger mapping units increase the ENL and SNR and provide more reliable coherence estimates, thereby enabling better predictions with less noisy inputs. Transition from 20~m to 60~m resolution resulted in an average increase in $R^2$ of approximately 0.1 and a reduction in RMSE by about 0.4--0.6~m across all input feature configurations. These improvements reflect the stabilizing effect of spatial aggregation, which reduces local speckle-induced variance and enhances the coherence of learned spatial patterns. Overall, these findings suggest that moderate spatial aggregation (40--60~m) provides the best trade-off between spatial detail and predictive stability for PALSAR-2 based deep learning retrievals.

At the finest resolution (20~m), the studied models achieved the highest spatial detail but also slightly larger noise-induced errors, with RMSE typically between 3.0--3.9~m and $R^2$ values around 0.63--0.67. Aggregation to 40~m led to modest improvements in accuracy for most configurations (RMSE $\approx$ 2.7--3.3~m; $R^2$ up to 0.73--0.76), suggesting that spatial averaging helps suppress speckle and local variability effects. Further coarsening to 60~m produced very limited additional gains, with RMSE as low as 1.9--2.1~m and $R^2$ values reaching up to 0.78--0.79 for the best-performing deep-learning models, indicating that likely 40m resolution can provide an optimal trade off between prediction accuracy and spatial detail.

Across all resolutions, both enhanced architectures— Nested U-Net and SeU-Net—outperformed the baseline U-Net. The Nested U-Net demonstrated more consistent performance across feature combinations and scales, reflecting the benefit of dense skip connections for multiscale feature integration. Meanwhile, SeU-Net often achieved slightly higher accuracy at coarser resolutions (particularly 60~m) and for complex input configurations, likely due to its channel-wise recalibration mechanism that adaptively weights informative features. The baseline U-Net consistently lagged behind these variants, though still exceeded the accuracy of conventional machine learning methods summarized in  Section \ref{subsec:comparisons_shallowML}.

Consequently, resolutions between 40~m and 60~m appeared to provide the best balance between spatial fidelity and predictive accuracy for PALSAR-2 based deep learning retrievals.

\subsection{Comparison with baseline ML methods} 
\label{subsec:comparisons_shallowML}

Traditional machine-learning methods exhibited predictable scaling trends: Random Forest (RF) and kNN generally outperformed MLR, with RF achieving RMSE $\approx$3.1~m at 20~m and $\approx$2.2~m at 60~m. However, examined UNet-family models consistently surpassed them by 0.3--0.5~m across resolutions, demonstrating superior capacity to generalize nonlinear dependencies among backscatter intensity, interferometric coherence, and geometry-derived features.

Visual inspection of Figure \ref{fig:patches} indicates that UNet models reproduce the reference spatial patterns more precisely than the baseline methods. The Nested U-Net and SeU-Net architectures capture both the fine-scale heterogeneity within forest stands and the broader gradients in canopy height across the scene. Their outputs show close correspondence with the reference, particularly in densely forested areas, where local maxima and low-height clearings are clearly delineated. In contrast, the baseline methods  exhibit stronger smoothing effects (especially MLR) and reduced dynamic range (all methods), leading to underestimation of tall canopies and less distinct stand boundaries. For Random Forests and kNN, some nonstationarity is evident in adjacent pixels even in homogeneous areas explained by pixel-level prediction approach.

The addition of interferometric coherence and polarimetric variables ($\gamma_{14}$, $\tau$, $\rho_\infty$) improves structural detail and contrast across all models, but the benefit is more apparent for the UNet model predictions. While Random Forest partly recovers large-scale gradients, the deep models produce maps whose textural and radiometric patterns align closely with the ALS based reference, allowing to produce spatially consistent height retrievals.

\subsection{Comparison with prior literature} 
\label{subsec:comparisons_prior}

Results reported in our paper for deep learning architectures appear more competitive when compared to recent state-of-the-art studies employing both traditional model-based and modern deep learning approaches across different sensors and forest biomes.

Model-based approaches using spaceborne L-band time series have reported RMSE values typically ranging from $\sim$ 4 to 6 m, depending on the study site and dataset. \cite{seppi2022} achieved RMSEs of 4--5 m from SAOCOM L-band InSAR coherence over managed forests in Argentina.
Using PALSAR-2 time series, \cite{Telli2025} showed that combining ALOS-2 backscatter and interferometric temporal coherence reduced RMSE to $\sim$4.2 m over mixed evergreen and deciduous forests, with class-specific errors between 2.0 and 6.2 m.
Comparable accuracies were obtained by \cite{Zhang2022}, who reported RMSEs of 4.4–4.8 m using single-baseline L-band PolInSAR inversion, and by \cite{lei2014} who achieved RMSE below 4~m using repeat-pass InSAR coherence over Maine, USA. Earlier airborne L-band PolInSAR work by \cite{praks2012} similarly reported RMSE $\approx$3.7~m using different versions of simplified LIDAR-DEM aided RVoG inversion. In a study over Traunstein site in Germany, a similar accuracy with RMSE of 3.71 m was achieved when using RVoG-model based retrievals with L-band PolInSAR images. 

Finally, a regionally focused study over Asturias region in Spain using multi-source Sentinel-1 and PALSAR-2 mosaic layers and optical Sentinel-2 images and wide range of machine-learning methods (RF, GPR, and kNN) obtained RMSE values of approximately 3--5~m for forest height mapping \citep{teijidomurias2025}. These results fall within the same accuracy range as the present study for pixel-level machine learning predictions, yet here the use of UNet models and interferometric features allowed to obtain superior results using PALSAR-2 time series in contrast to multi-source dataset used in prior paper. 

Overall, our studied models, especially the attention-augmented SeU-Net and Nested UNet variants, achieved lower RMSE values than physics model-based and classical machine-learning approaches relying on similar L-band data, or even more advanced PolInSAR datasets or multi-source SAR and optical imagery, confirming the potential of the proposed deep learning framework for tree height estimation from PALSAR-2 time series.

\subsection{Model applicability and limitations}
Described approach is suitable for regional mapping under similar weather and seasonal conditions when there is a representative set of reference ALS data (hundreds of image patches) available for end-to-end training. 

Similar performance can be expected in adjacent regions with comparable forest composition. However, applications in remote areas or different forest biomes would require significant model fine-tuning, such as domain adaptation or model transfer, supported by at least a small sample of reference ALS patches or a representative set of forest plots. The potential of pretrained model transfer in the forest context has been studied to a limited extent under simplified scenarios \citep{ge_transfer2023, kuzu2024}. Further research is needed to investigate more complex cases, as radar scattering signatures depend not only on forest structure but also on seasonal and weather conditions influencing forest transmissivity and ground conditions (e.g., rainfall, soil moisture, or frozen conditions), as well as on imaging geometry and topography.  

\section{Conclusions and Outlook}\label{sec:conclusion}

This study presented a deep learning framework to estimate tree height from time series of L-band ALOS-2 PALSAR-2 interferometric acquisitions. Three DL models from the U-Net family, i.e., U-Net, Nested U-Net, and SeU-Net, were used to asses the impact of architecture design on retrieval accuracy. Nested skip connections as well as channel-wise recalibration significantly improved performance compared to basic U-Net. Among the tested models, Nested U-Net and SeU-Net achieved comparable accuracy, with the former yielding the best overall results. 

Several combinations of SAR and InSAR input data were tested to evaluate the contribution of multi-polarization and multi-temporal features.  Integrating temporal coherence with backscatter significantly reduced estimation errors, achieving the best results when the full temporal coherence stack was exploited. Similar accuracy was obtained when using model-derived parameters ($\tau$, $\rho_\infty$), demonstrating that these features effectively capture the temporal and spatial sensitivity of coherence to forest structure while reducing input dimensionality. The inclusion of geometry layers, such as incidence angle and DEM, further enhanced performance by allowing the network to better account for topography-induced variations in backscatter and coherence. The best accuracy at 20 m resolution was achieved using the full feature stack, with an RMSE of approximately 2.8~m.

The comparison across mapping scales reported higher estimation accuracies at 40 m and 60 m, achieving RMSE $\sim$ 1.95 m at coarser resolution. The achieved accuracies outperform those reported in recent studies based on model-based and traditional machine learning approaches, confirming the strong potential of deep learning for SAR-based forest structure retrieval.

Future developments will include the application of the proposed framework to upcoming NISAR data, enabling large-scale assessments across diverse forest biomes. Additional efforts will focus on integrating complementary datasets, such as optical and C-band SAR observations, to further enhance retrieval accuracy and generalization. Finally, testing alternative deep learning architectures and incorporating explainable AI techniques will be essential to improve model interpretability and advance the physical understanding of SAR-based forest parameter retrieval.

\section*{Acknowledgments}
The authors would like to thank Japan Aerospace Exploration Agency - JAXA for providing ALOS-2 data and NASA JPL for providing the ISCE3 software.

\bibliographystyle{elsarticle-harv} 
\bibliography{lib,telli}

\end{document}